\documentclass{elsart}
\usepackage{amssymb}
\usepackage{graphicx}

\journal{Physica A}

\begin{document}

\begin{frontmatter}

\title{Criticality in the randomness-induced second-order phase transition of the triangular
Ising antiferromagnet with nearest- and next-nearest-neighbor
interactions}

\author{N. G. Fytas\corauthref{cor1}}, \corauth[cor1]{Corresponding author.}
\ead{nfytas@phys.uoa.gr} \author{A. Malakis}

\address{Department of Physics, Section of Solid State Physics, University of Athens, Panepistimiopolis, GR 15784
Zografos, Athens, Greece}

\begin{abstract}
Using a Wang-Landau entropic sampling scheme, we investigate the
effects of quenched bond randomness on a particular case of a
triangular Ising model with nearest- ($J_{nn}$) and
next-nearest-neighbor ($J_{nnn}$) antiferromagnetic interactions.
We consider the case $R=J_{nnn}/J_{nn}=1$, for which the pure
model is known to have a columnar ground state where rows of
nearest-neighbor spins up and down alternate and undergoes a weak
first-order phase transition from the ordered to the paramagnetic
state. With the introduction of quenched bond randomness we
observe the effects signaling the expected conversion of the
first-order phase transition to a second-order phase transition
and using the Lee-Kosterlitz method, we quantitatively verify this
conversion. The emerging, under random bonds, continuous
transition shows a strongly saturating specific heat behavior,
corresponding to a negative exponent $\alpha$, and belongs to a
new distinctive universality class with $\nu=1.135(11)$,
$\gamma/\nu=1.744(9)$, and $\beta/\nu=0.124(8)$. Thus, our results
for the critical exponents support an extensive but weak
universality and the emerged continuous transition has the same
magnetic critical exponent (but a different thermal critical
exponent) as a wide variety of two-dimensional (2d) systems
without and with quenched disorder.
\end{abstract}

\date{\today}


\end{frontmatter}

\section{Introduction}
\label{sec:1}

Understanding the role played by impurities on the nature of phase
transitions is of major importance, both from experimental and
theoretical perspectives. It has been known that quenched bond
randomness may or may not modify the critical exponents of
second-order phase transitions, based on the Harris
criterion~\cite{harris-74,berker-90}. It was more recently
established that quenched bond randomness always affects
first-order phase transitions by conversion to second-order phase
transitions, for infinitesimal randomness in
$d=2$~\cite{aizenman-89,hui-89} and after a threshold amount of
randomness in $d>2$~\cite{hui-89}, as also inferred by general
arguments~\cite{berker-93}.

In the last $15$ years, softening effects on first-order
transitions have been studied and confirmed in several
investigations~\cite{uzelac-95,ballesteros-00,chatelain-01,fernandez-08,falicov-96,chen-92,kardar-95,cardy-97,chatelain-98,paredes-99,dotsenko-95,picco-97,cardy-99,ludwig-87}.
In particular in 2d the behavior of the random-bond ($q$-state)
Potts model (RBPM) has attracted special attention, since by
varying $q$ one can observe the expected softening effects on both
second- and first-order ($q>4$) phase transitions. For this model
the extensive numerical study of Cardy and
Jacobsen~\cite{cardy-97} clearly showed the existence of a new
critical behavior for each value of $q$, independently of the
disorder. This fact was illustrated by producing a convincing
continuous behavior for the magnetic exponent $\beta/\nu$ as
function of $q$, exhibiting no singularity at $q=4$. Their results
are in excellent agreement (up to $q=4$) with the theoretical
predictions, in the vicinity of $q=2$, originally of
Ludwig~\cite{ludwig-87} and improved by Dotsenko et
al.~\cite{dotsenko-95}. Thus, the study of Cardy and
Jacobsen~\cite{cardy-97} has resolved and critically discussed
the controversy related to the early simplifying prediction from
the numerical simulations on the $q=8$ RBPM, that the emerging
second-order phase transitions may exhibit critical exponents
consistent with those of the pure Ising
($q=2$) model~\cite{chen-92,kardar-95}.

The above mentioned simplifying picture, that in a variety of
situations the universality class of random-bond models is that of
the Ising model~\cite{chen-92,kardar-95,wiseman-95} cannot be
also valid in systems with competing interactions, where a strong
saturating behavior of the specific heat has been observed
recently~\cite{fytas-08a,malakis-09}. In the present paper, we
shall study a further such novel second-order phase transition
induced by bond randomness. In the corresponding pure generalized
Ising system, the competition of the microscopic nearest- and
next-nearest-neighbor exchange interactions gives rise to a weak
first-order phase transition. We find now a rather delicate
situation, where the introduction of a rather weak bond randomness
on the also weak first-order transition of the pure system
produces a dramatic saturating behavior on the specific heat and a
new critical behavior.

\begin{figure}[htbp]
\includegraphics*[width=12 cm]{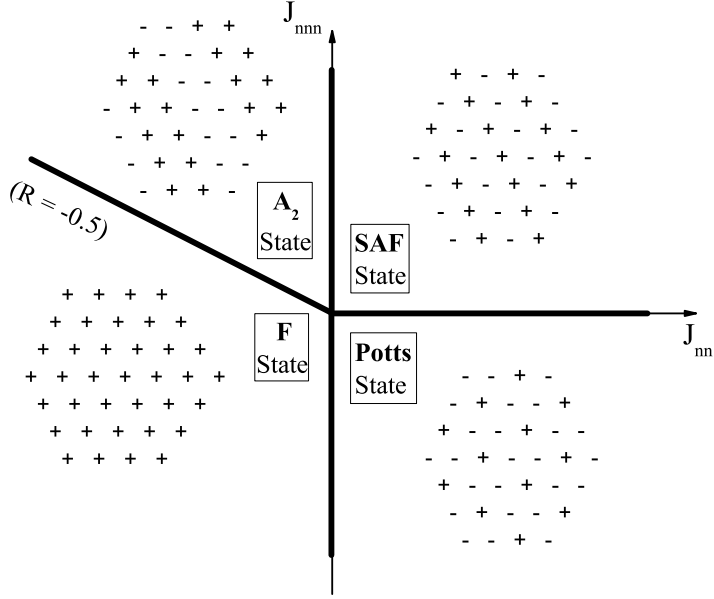}
\caption{\label{fig:1}$T=0$ phase diagram of the triangular Ising
model with nearest- and next-nearest-neighbor interactions.}
\end{figure}
The generalized (pure) Ising Hamiltonian is described by the
following
\begin{equation}
\label{eq:1} H_{p}=J_{nn}\sum_{\langle
i,j\rangle}S_{i}S_{j}+J_{nnn}\sum_{(i,j)}S_{i}S_{j},
\end{equation}
and the Ising spin system is embedded on a 2d triangular lattice.
In terms of the exchange interactions, the pure spin system on the
triangular lattice obeys, depending on the interactions, a variety
of interesting orderings~\cite{metcalf-74,tanaka-75}, as
illustrated by the ground-state phase diagram in
figure~\ref{fig:1}. For the case of positive (antiferromagnetic) nearest- $J_{nn}$ and
next-nearest-neighbor $J_{nnn}$ competing
interactions, the illustrated layered ground state corresponds to
a six-fold degenerate arrangement in which ferromagnetic lines
alternate with opposite oriented spins in the three lattice
directions~\cite{metcalf-74,tanaka-75}. We shall refer to the
corresponding ordered phase as the superantiferromagnetic (SAF)
phase and the corresponding triangular (Tr) model as the TrSAFM.
For this case, Rastelli et al.~\cite{rastelli-05} have shown via
Monte Carlo (MC) simulations that the system undergoes a
first-order phase transition from the SAF phase to a
high-temperature paramagnetic phase by studying the ratio of
interactions $R=J_{nnn}/J_{nn}=0.1$, $0.5$, and $1$. For the
particular case $R=1$ our study~\cite{malakis-07} verified and
improved the results of Rastelli et al.~\cite{rastelli-05},
showing that the first-order transition has characteristics that
lie between those of the $5$- and $6$-states Potts model.

We now consider here the random-bond version of this last case by
introducing the well-known bimodal distribution of bond disorder,
applied on both nearest- and next-nearest-neighbor spins $i$ and
$j$
\begin{equation}
\label{eq:2}
P(J_{ij})=\frac{1}{2}[\delta(J_{ij}-J_{1})+\delta(J_{ij}-J_{2})];\;\;
\frac{J_{1}+J_{2}}{2}=1,
\end{equation}
where the new ratio $r=J_{2}/J_{1}$ of the weak to the strong
bonds denotes the disorder strength and we fix
$2k_{B}/(J_{1}+J_{2})=1$ to set the temperature scale. Thus, using
the above distribution into Hamiltonian~(\ref{eq:1}), the
resulting random-bond version (RBTrSAFM) reads now as
\begin{equation}
\label{eq:3}
H=\sum_{<i,j>}J_{ij}S_{i}S_{j}+\sum_{(i,j)}J_{ij}S_{i}S_{j}.
\end{equation}
In the present study, we will restrict ourselves only on the
rather weak value of disorder strength $r=0.9/1.1$, a value,
however, that will produce a dramatic saturation of the originally
pure model's $L^{d}$ - divergence of the specific heat.

The rest of the paper is organized as follows: In the next Section
we outline our numerical scheme and in Section~\ref{sec:3} we
present and discuss our results on the RBTrSAFM. Specifically, in
Section~\ref{sec:3a} we illustrate the conversion, under the
presence of quenched bond randomness, of the first-order
transition of the pure model to second-order and using the
Lee-Kosterlitz method~\cite{lee-90} we quantitatively verify this
conversion. We continue in Section~\ref{sec:3b} by performing a
finite-size scaling (FSS) analysis on our data in order to
estimate the critical exponents describing the induced continuous
transition. Finally, we summarize our conclusions in
Section~\ref{sec:4}.

\section{Entropic simulation scheme} \label{sec:2}

Resorting to large scale MC simulations is often necessary and
useful~\cite{selke-94}, especially for the study of the critical
behavior of disordered systems. It is also
well-known~\cite{newman-99} that for such complex systems
traditional methods become very inefficient and that in the last
few years several sophisticated algorithms, some of them based on
entropic iterative schemes, have been proven to be very effective.
The present numerical study of the RBTrSAFM will be carried out by
applying our recent and efficient entropic
scheme~\cite{fytas-08a,malakis-04,fytas-08b}. In this approach we
follow a two-stage strategy of a restricted entropic sampling,
which is described in our study of random-bond Ising models (RBIM)
in 2d~\cite{fytas-08a} and is very similar to the one applied also
in our numerical approach to the 3d random-field Ising model
(RFIM)~\cite{fytas-08b}. In these papers, we have presented in
detail the various sophisticated routes used for the restriction
of the energy subspace and the implementation of the Wang-Landau
(WL) algorithm~\cite{wang-01}. Therefore, only a brief outline
will be presented bellow.

The identification of the appropriate energy subspace
$(E_{1},E_{2})$ for the entropic sampling of each random-bond
realization is carried out by applying our critical minimum energy
subspace restriction~\cite{malakis-04} and taking the union
subspace at both pseudocritical temperatures of the specific heat
and magnetic susceptibility. This subspace, extended by $10\%$ from both
low- and high-energy sides, is sufficient for an accurate
estimation of all finite-size anomalies. Following
references~\cite{fytas-08a,fytas-08b}, the identification of the
appropriate energy subspace is carried out in the first
multi-range (multi-R) WL stage in a wide energy subspace. The WL
refinement levels ($G(E)\rightarrow f G(E)$, where $G(E)$ is the
density of states (DOS); for more details see
references~\cite{fytas-08a,fytas-08b}) used in this stage
($j=1,\ldots,j_{i}; f_{j+1}=\sqrt{f_{j}}$) were $j_{i}=19$ for $L<
100$ and $j_{i}=20$ for $L\geq 100$. The same process was repeated
several times, typically $\sim 5$ times, in the newly identified
restricted energy subspace and the average DOS was determined. From our
experience, this repeated application of the first multi-R WL approach greatly improves
accuracy and then the resulting accurate DOS is used for a final
redefinition of the restricted subspace, in which the final
second stage is then applied.

In the second entropic stage we accumulate new and more accurate
$(E,M)$ histogram data, by implementing now the refinement WL
levels $j=j_{i}+1,\ldots,j_{i}+4$, improving also eventually the final DOS. Now, the
implementation can be carried out in an one-range (one-R) or in a multi-R fashion.
However, our comparative study of the first-order transition
features of the 3d RFIM~\cite{fytas-08b} has revealed the need for
a careful one-R implementation of the final stage,
in order to probe correctly the possible double-peaked (dp)
structure of the energy probability density function (PDF). Thus,
having to do with the RBTrSAFM and possible remaining
characteristics of the dp structure of the pure model, we carried
out the final stage of our approach in various ways and verified
that both a demanding multi-R (with larger energy pieces) and also
an one-R approach were very successful and gave comparable
results. Furthermore, we carried out an alternative and very
simple one-R approach, in which the WL modification factor was
adjusted according to the rule $\ln f\sim t^{-1}$ proposed
recently by Belardinelli and Pereyra~\cite{belardinelli-07}. Our
comparative tests showed that this last approach gave very good
results, for all disorder realizations, and was therefore used for
most of our subsequent simulations.

\begin{figure}[htbp]
\includegraphics*[width=12 cm]{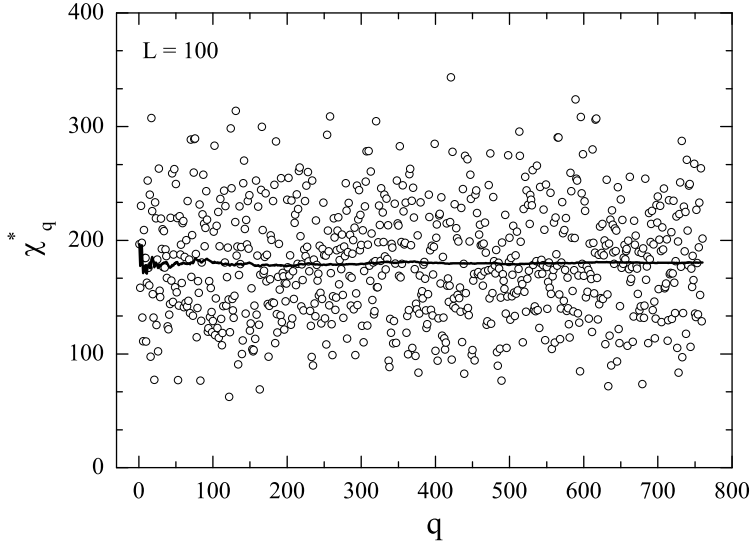}
\caption{\label{fig:2}Disorder distribution of the susceptibility
maxima of a lattice of linear size $L=100$. The running average
over the samples is shown by the thick solid line.}
\end{figure}

Thus, using this scheme, we performed extensive simulations for
several lattice sizes in the range $L=20-160$, over large
ensembles $\{1,\ldots,q,\ldots,Q\}$ of random realizations -
$Q_{L\leq 40}=256$, $Q_{L\leq 80}=512$, and $Q_{L\geq 100}=768$.
It is well-known that, extensive disorder averaging is necessary
when studying random systems, especially near ex-first-order
transitions, where, as was shown by Fisher~\cite{fisher-95},
extremely broad distributions are expected leading to a strong
violation of self-averaging~\cite{aharony-96,wiseman-98}.
Figure~\ref{fig:2} presents evidence that the above number of
random realizations is sufficient in order to obtain the true
average behavior and not a typical one. In particular, we plot in
this figure (for $L=100$) the disorder distribution of the
susceptibility maxima $\chi_{q}^{\ast}$ and the corresponding
running average, i.e. a series of averages of different subsets of
the full data set - each of which is the average of the
corresponding subset of a larger set of data points, over the
samples for the simulated ensemble of $Q=768$ disorder
realizations. A first striking observation from this figure is the
existence of very large variance of the values of
$\chi_{q}^{\ast}$, indicating the expected violation of
self-averaging for this quantity. This figure illustrates that the
simulated number of random realizations is sufficient in order to
probe correctly the average behavior of the system, since already
for $Q\approx 400$ the average value of $\chi^{\ast}_{q}$ appears
quite stable.

Closely related to the above issue of self-averaging in disordered
systems is the manner of averaging over the disorder. This
non-trivial process may be performed in two distinct ways when identifying the finite-size
anomalies, such as the peaks of the magnetic susceptibility. The first way corresponds to
the average over disorder configurations
($[\ldots]_{av}$) and then taking the maxima
($[\ldots]^{\ast}_{av}$), or taking the maxima in each individual
configuration first, and then taking the average
($[\ldots^{\ast}]_{av}$). In the present paper we have undertaken
our FSS analysis using both ways of averaging and have found
comparable results for the values of the critical exponents, as
will be discussed in more detail below. Closing this Section, let
us comment on the statistical errors of our numerical data to be
presented in the next section. The statistical errors of our WL
scheme, on the observed averaged behavior, were found to be of
relatively small magnitude (of the order of the symbol sizes) when
compared to the relevant disorder-sampling errors due to the
finite number of simulated realizations. Thus, the error bars
shown in our figures below, used in the corresponding fitting
attempts, reflect the disorder-sampling errors and have been
estimated using groups of $32$ or $64$ realizations via the
jackknife method~\cite{newman-99}.

\section{Phase transition of the triangular random-bond SAF model}
\label{sec:3}

\subsection{Nature of the transition}
\label{sec:3a}

In the first part of our study we determine the order of the phase
transition. This has traditionally been rather difficult for
systems showing weak first-order transitions, but finite-size
effects at first-order transitions are now much better
understood~\cite{pleimling-01,binder-03}. As it is well known from
the existing theories of first-order transitions, all finite-size
contributions enter in the scaling equations in powers of the
system size $L^{d}$~\cite{fisher-82}. This holds for the general
shift behavior (for various pseudotransition temperatures) and
also for the FSS behavior of the peaks of various energy cumulants
and of the magnetic susceptibility. It is also well known that the
dp structure of the energy PDF, $P_{L}(e)$, where $e=H/L^{d}$, is
signaling the emergence of the expected two delta-peak behavior in
the thermodynamic limit, for a genuine first-order phase
transition~\cite{binder-84,binder-87}, and with increasing lattice
size the barrier between the two peaks should steadily increase.

\begin{figure}[htbp]
\includegraphics*[width=14 cm]{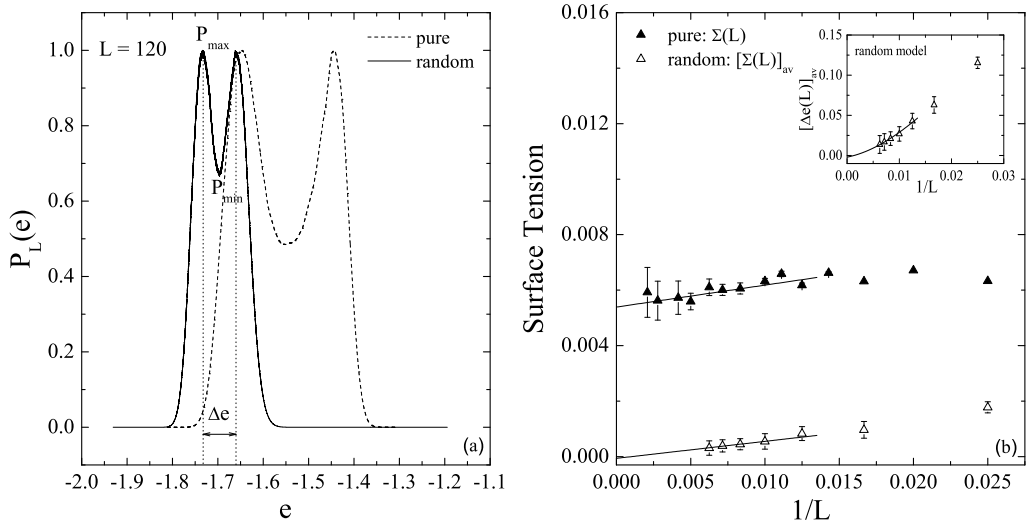}
\caption{\label{fig:3}(a) Softening effects induced by bond
randomness on the dp structure of the energy PDF of the TrSAFM.
(b) Limiting behavior of the surface tension of the pure
($\Sigma(L)$: filled triangles) and the RBTrSAFM
($[\Sigma(L)]_{av}$: open triangles). The inset shows the limiting
behavior of the width $[\Delta e(L)]_{av}$ for the random model.}
\end{figure}

According to the arguments of Lee and Kosterlitz~\cite{lee-90} the
surface tension $\Sigma(L)=\Delta
F(L)/L^{d-1}=[k_{B}T\ln{(P_{max}/P_{min})}]/L$, where $P_{max}$
and $P_{min}$ are the maximum and minimum energy PDF values at the
temperature $T_{h}$ where the two peaks are of equal height (see
figure~\ref{fig:3}), should tend to a non-zero value.
Figure~\ref{fig:3}(a) shows this pronounced dp structure of the
energy PDF of the pure and the random model (for a particular
realization of bond disorder) at the corresponding temperatures
where the two peaks are of equal height, for a lattice size
$L=120$. It appears that the dp structure of the PDF is partly
maintained for the RBTrSAFM and relatively small lattices, but it
is clear that the introduction of randomness has induced
significant softening effects. With the help of
figure~\ref{fig:3}(a) we may define the surface tension
$\Sigma(L)$ and also the width $\Delta e(L)$ that represents, in
the limit $L\rightarrow \infty$, the latent heat of the
transition, in the case of a first-order phase transition.

Following Chen et al.~\cite{chen-92} we define the disorder
averaged surface tension
\begin{equation}
\label{eq:4}
[\Sigma(L)]_{av}=\frac{1}{Q}\sum_{q=1}^{Q}\Sigma_{i}(L)
\end{equation}
and the corresponding width of the transition
\begin{equation}
\label{eq:5} [\Delta e(L)]_{av}=\frac{1}{Q}\sum_{q=1}^{Q}\Delta
e_{i}(L),
\end{equation}
which are for the present case the proper measures in order to
apply the Lee-Kosterlitz~\cite{lee-90} FSS argument.
Figure~\ref{fig:3}(b) shows the limiting behavior of these two
quantities, $[\Sigma(L)]_{av}$ in the main panel and $[\Delta
e(L)]_{av}$ in the corresponding inset, for the RBTrSAFM,
including also, in the main panel, the corresponding surface
tension points of the pure model~\cite{malakis-07}. The solid
lines in the main panel are simple linear fittings for the larger
lattice sizes, indicating for the random model a zero value for
$[\Sigma(L)]_{av}$ in the limit $L\rightarrow \infty$ and
therefore a clear asymptotic conversion of the originally
first-order transition to a second-order transition transition
under the presence of the quenched random-bond
distribution~(\ref{eq:2}), even for the weak disorder strength
$r=0.9/1.1$. The solid line in the corresponding inset shows a
second-order polynomial fitting for the data of $[\Delta
e(L)]_{av}$ which appears to be the best fitting describing the
approach of $[\Delta e(L)]_{av}$ to zero in the asymptotic limit.

\subsection{Finite-size scaling - Critical exponents}
\label{sec:3b}

\begin{figure}[htbp]
\includegraphics*[width=14 cm]{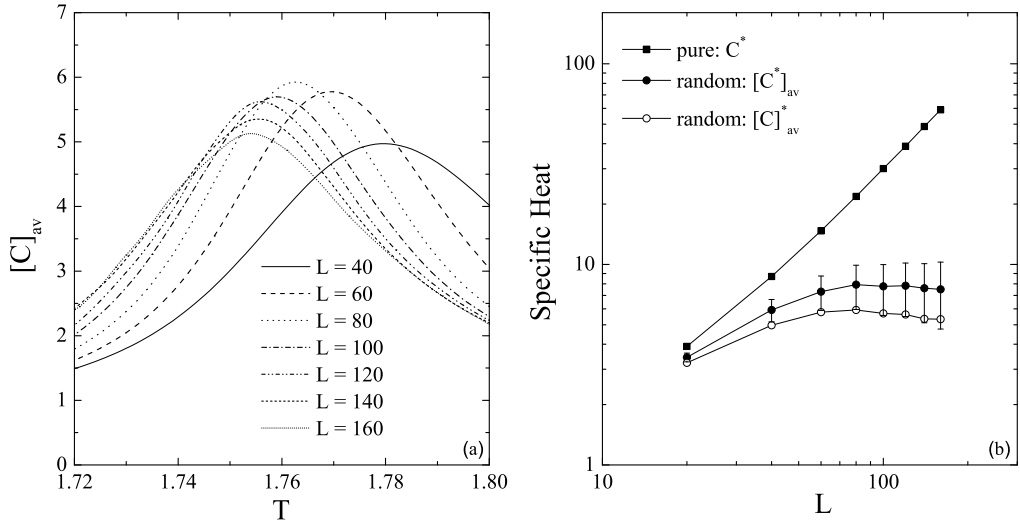}
\caption{\label{fig:4}(a) Average specific heat curves as a
function of temperature for several lattice sizes in the range
$L=40-160$. (b) Size dependence of the maxima of the specific heat
of the pure (filled squares) and random-bond TrSAFM (filled and open
circles) in a log-log scale. The error bars reflect the
sample-to-sample fluctuations of $[C^{\ast}]_{av}$.}
\end{figure}

We proceed now with a quantitative characterization of the induced
second-order phase transition of the model, by providing estimates
for the critical exponents. Let us start the presentation of our
results with the most striking effect of the bond randomness on
the specific heat of the TrSAFM. In figure~\ref{fig:4}(a) we
present the average specific heat curves $[C]_{av}$ as a function
of temperature for several lattice sizes in the range $L=40-160$.
The typical average curve is a smooth function of temperature, as
expected, and its maximum $[C]_{av}^{\ast}$ increases for sizes
$L\leq 80$ and clearly decreases for sizes $L>100$. The remaining
dp structure of the energy PDF [see figure~\ref{fig:3}(a)]
becomes less important for large lattices (the two peaks come
closer) and the limiting height will be finite, a behavior in
agreement with an expected asymptotic cusp and a negative exponent
$\alpha$. In figure~\ref{fig:4}(b) we now contrast the size
dependence of the specific heat maxima of both the pure and
random-bond TrSAFM. For the random case we present two data set
points corresponding to the two averaging processes, i.e. the
upper curve corresponds to the average of the individual maxima
$[C^{\ast}]_{av}$ and the error bars shown are the
sample-to-sample fluctuations of this quantity, whereas the lower
curve corresponds to the maxima of the averaged curve
$[C]_{av}^{\ast}$. From figure~\ref{fig:4} the suppression of the
specific heat maxima is very clear and the asymptotic saturation
of the specific heat seems to be unquestionable. Noteworthy here
that, this behavior is similar to the one found in our recent
studies of random-bond systems with competing
interactions~\cite{fytas-08a,malakis-09}, but very different from
that observed in the cases of the
RBIM~\cite{fytas-08a,dotsenko-81,shalaev-84,shankar-87,ludwig-87b,wang-90,reis-96,ballesteros-97,selke-98,mazzeo-99}
and the $q=8$ RBPM~\cite{chen-92}.

\begin{figure}[htbp]
\includegraphics*[width=14 cm]{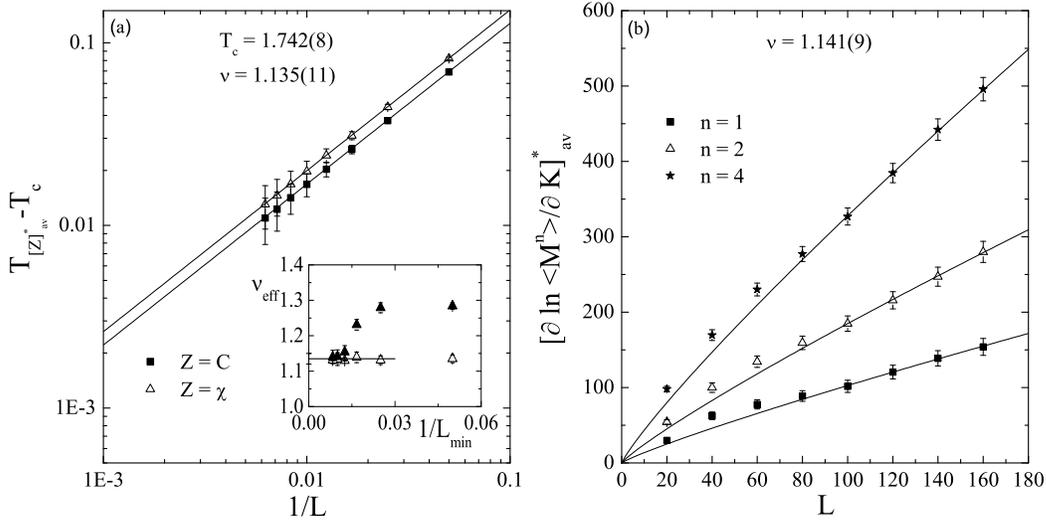}
\caption{\label{fig:5}(a) Simultaneous fitting
(equation~(\ref{eq:6}) in the range $L=20-160$) of two
pseudocritical temperatures, $Z=C$ (filled squares) and $Z=\chi$
(open triangles), giving the values $T_{c}=1.742(8)$ and
$\nu=1.135(11)$. The data are shown in a double logarithmic scale.
(b) Simultaneous fitting (simple power-law for $L\geq 100$) of the
average logarithmic derivatives for $n=1,2$, and $4$ giving the
value $\nu=1.141(9)$. The inset presents values of effective
exponents $\nu_{eff}$ obtained from the data of the pseudocritical
temperatures (open reversed triangles) and the logarithmic
derivatives (open triangles) corresponding to panels (a) and (b).
The solid line in the inset marks the proposed estimate
$\nu=1.135(11)$.}
\end{figure}

The above remarks about the behavior of the specific heat of the
RBTrSAFM implicitly suggest a new critical behavior. In order to
estimate the critical exponents describing this disorder-induced
continuous transition, we will now follow the standard FSS methods
applying several alternatives for their estimation and use an
appropriate procedure to evaluate the stability of our estimates.
Such a stability test is useful in cases where one may expect
finite-size problems. Such well-known problems may be due to
possible cross-over effects, as those discussed by
Picco~\cite{picco-98} for the case of the RBPM. Also, they may
appear in the presence of logarithmic corrections and in the past
produced erroneous exponents, as explained by Ballesteros et
al.~\cite{ballesteros-97} for the case of the 2d site-diluted
Ising model. In particular, the determination of the correlation
length exponent will be carried out by two alternatives. In the
first, we observe and analyze the shift behavior of two
pseudocritical temperatures corresponding to the maxima of the
average specific heat and magnetic susceptibility
($T_{[Z]^{\ast}_{av}}$), according to the usual shift relation
\begin{equation}
\label{eq:6} T_{[Z]^{\ast}_{av}}=T_{c}+bL^{-1/\nu}.
\end{equation}
Furthermore, we report here the analogous behavior corresponding
to the pseudocritical temperatures $[T_{Z}^{\ast}]_{av}$ of the
individual maxima. In the second alternative we study the
divergences of the logarithmic derivatives of some powers of the
order parameter with respect to the inverse temperature
$K=1/T$~\cite{chen-92}.

The shift behavior of the pseudocritical temperatures
corresponding to the maxima of the average curves is shown in the
main panel of figure~\ref{fig:5}(a). The solid lines in this panel
are a simultaneous fitting using data from all lattice sizes
($L=20-160$), giving a convincing behavior and a value
$T_{c}=1.742(8)$ for the critical temperature of the random
system, well below the corresponding transition temperature
$T^{\ast}_{p}=1.8084$ of the pure system~\cite{malakis-07}. The
estimate of the correlation length exponent from the fitting in
figure~\ref{fig:5}(a) is $\nu=1.135(11)$, in agreement with the
inequality $\nu\geq 2/d$ for disordered systems of Chayes et
al.~\cite{chayes-86}. A similar simultaneous fitting attempt was
performed, but not shown for brevity, to the numerical data for
the pseudocritical temperatures $[T_{Z}^{\ast}]_{av}$ that
corresponds to the averaging process of the individual maxima. The
corresponding estimates are: $T_{c}=1.736(9)$ and $\nu=1.139(12)$.

The second alternative estimation of $\nu$ is carried out by
analyzing the divergency of the logarithmic derivatives of the
order parameter defined as
\begin{equation}
\label{eq:7} \frac{\partial \ln \langle M^{n}\rangle}{\partial
K}=\frac{\langle M^{n}E\rangle}{\langle M^{n}\rangle}-\langle
E\rangle.
\end{equation}
As is well-known~\cite{chen-92}, the corresponding maxima scale
with the system size as $\sim L^{1/\nu}$. In panel (b) of
figure~\ref{fig:5} we illustrate the size dependence of the first-
(filled squares), second- (open triangles), and fourth-order
(filled stars) maxima of the average logarithmic derivatives. Now,
the solid lines show a simultaneous power-law fitting using only
the larger lattice sizes ($L\geq 100$) and provide an estimate
$\nu=1.141(9)$.

Finally, the inset of figure~\ref{fig:5}(a) illustrates our method
to evaluate and discuss the stability of the estimation for the
exponent $\nu$. It shows values of effective exponents
($\nu_{eff}$) determined by imposing a lower cutoff $(L_{min})$
and applying simultaneous fittings in windows $(L_{min}-L_{max})$,
where $L_{max}=160$ and $L_{min}=20,40,60,80,100$, and $120$ as a
function of $1/L_{min}$. The effective estimates, from the
pseudocritical temperatures (inverse open triangles) are very
stable around the value $1.135(11)$ and their behavior is a clear
evidence in favor of the suggested new critical behavior. Their
values don't show any trend or tendency towards the value $\nu=1$
of the Ising universality class. On the other hand, the effective
estimates from the logarithmic derivatives show a rather strong
cross-over effect for small values of the lower cutoff. For larger
values a trend for settlement to the value $1.135$ (marked by the
solid line) may be observed as $L_{min}\rightarrow 120$. These
observations explain also the reasons why we have used in the main
panels (a) and (b) of figure~\ref{fig:5} different windows for the
proposed estimations. In conclusion, we feel that, the emerging
from this figure picture rules out the possibility that, our main
suggestion, of a new critical behavior, is a result of finite-size
or cross-over effects. Of course, there is always some danger
that, strong effects, persisting to very large lattices, may
produce erroneous results and critical exponents. This fact has
been pointed out in several papers dealing with disordered
systems~\cite{ballesteros-97,picco-98,andreichenko-90}.

\begin{figure}[htbp]
\includegraphics*[width=14 cm]{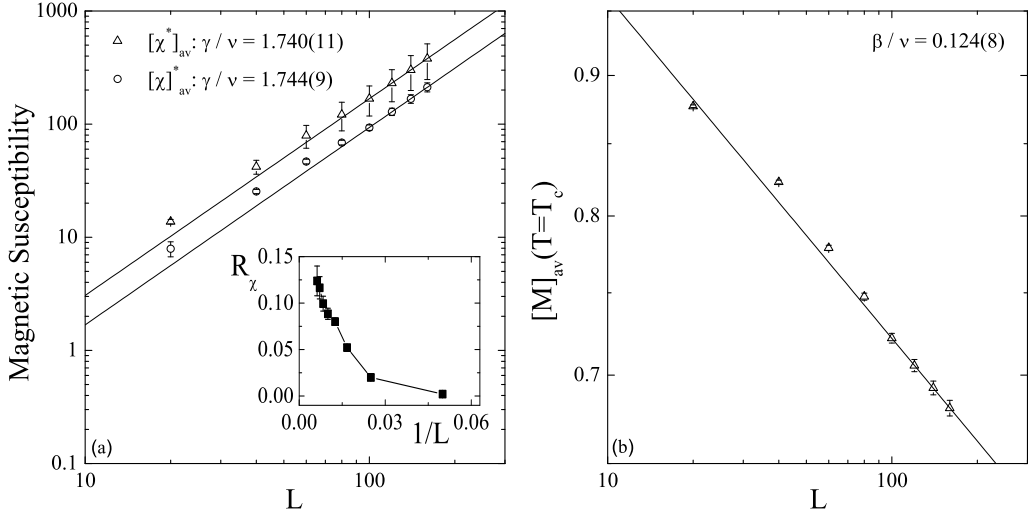}
\caption{\label{fig:6}(a) Log-log plots of the size dependence of
the maxima of the average magnetic susceptibility
$[\chi]_{av}^{\ast}$ and the average of the individual maxima
$[\chi^{\ast}]_{av}$. The inset shows the limiting behavior of the
ratio $R_{[\chi^{\ast}]_{av}}$ defined in the text. (b) Estimation
of the magnetic exponent ratio $\beta/\nu$ from the average
magnetization at the estimated critical temperature. Linear fittings
are applied only on the data of sizes $L\geq 100$.}
\end{figure}

We turn now to the estimation of the magnetic exponents of the
RBTrSAFM. In figure~\ref{fig:6}(a) we present the FSS behavior of
the maxima of the average magnetic susceptibility
$[\chi]_{av}^{\ast}$ and also the average of the individual maxima
$[\chi^{\ast}]_{av}$. For the average $[\chi]_{av}^{\ast}$ the
error bars indicate the statistical errors due to the finite
number of the realizations, as discussed in Section~\ref{sec:2}.
In view of this discussion, the relatively small error bars
reflect the sufficiency of the disorder-averaging process and also
the accuracy of our two-stage implementation of the WL scheme. For
the average $[\chi^{\ast}]_{av}$ the errors bars shown reflect now
the relatively very large sample-to-sample fluctuations. Using
these sample-to-sample fluctuations, we construct the ratio
$R_{\chi}=V_{\chi}/[\chi]_{av}^{2}=([\chi^{2}]_{av}-[\chi]_{av}^{2})/[\chi]_{av}^{2}$
and plot it as a function of the inverse linear size, as shown in
the inset of figure~\ref{fig:6}(a). This ratio is a well-known
measure of self-averaging~\cite{aharony-96,wiseman-98} of the
corresponding physical quantity and clearly for the present model
the limiting value of $R_{\chi}$ is non-zero, indicating a strong
violation of self-averaging, as mentioned already in
Section~\ref{sec:2}. Following our earlier practice (panel (b) of
figure~\ref{fig:5}) for the logarithmic derivatives, we present
now by solid lines in figure~\ref{fig:6}(a) linear fittings using only
the larger lattice sizes ($L\geq 100$), giving the estimates
$1.744(9)$ and $1.740(11)$ for the ratio $\gamma/\nu$.

Finally, in figure~\ref{fig:6}(b), also in a double logarithmic
scale, we plot the data of the average order parameter at the
estimated critical temperature $T_{c}=1.742$ of the model,
$[M]_{av}(T=T_{c})$. The solid line is a linear fitting for the larger
lattice sizes giving the value $0.124(8)$ for the magnetic
exponent ratio $\beta/\nu$. The high accuracy of our estimate for
the critical temperature, as well as the sufficiency of the
performed disorder-averaging are again reflected in the relatively
small error bars, especially for the larger sizes where as may be
seen the behavior appears to be quite smooth. The above
estimations strongly indicate that the ratios $\gamma/\nu$ and
$\beta/\nu$ of the RBTrSAFM may share the values of the pure 2d
Ising model. This property is supporting an extensive version of
the generalized statement~\cite{kim-94,kim-96} of weak
universality~\cite{suzuki-74,gunton-75} and appears to be also
obeyed in some cases of disordered models, including the
corresponding random-bond version of the square SAF
model~\cite{fytas-08a}, as well as the case of the induced
second-order phase transition of the originally first-order
transition of the 2d Blume-Capel (BC) model~\cite{malakis-09}.
However, for the case of the RBIM the property of strong
universality (logarithmic corrections) applies, as clearly shown
in references~\cite{dotsenko-81,shalaev-84,shankar-87,ludwig-87b,wang-90,reis-96,ballesteros-97,selke-98,mazzeo-99}.
Furthermore, as pointed out in the introduction for the
corresponding induced second-order phase transitions in the case
of the RBPM, this property is ruled out by the study of Cardy and
Jacobsen~\cite{cardy-97}, since in this paper a clear $q$
dependence has been shown for the magnetic exponent $\beta/\nu$.

We close this Section by determining the ratio $\alpha/\nu$ via
the Rushbrooke relation:
$\alpha/\nu=2/\nu-2\beta/\nu-\gamma/\nu=-0.23(4)$. As expected,
this negative value reflects the saturation property of the
specific heat, observed here but also in our similar recent
studies~\cite{fytas-08a,malakis-09}. Of course, the estimated
exponents satisfy here also hyperscaling, although attention
should be drawn for possible violations of hyperscaling in random
systems~\cite{schwartz-91}. From the above discussion and the
corresponding referenced literature we may conclude that, there
are still interesting and not settled questions for the origin of
the universality properties of the disorder-induced continuous
phase transitions and the problem remains of current interest.

\section{Conclusions}
\label{sec:4}

We have illustrated and quantitatively verified, using the
Lee-Kosterlitz method, that even a weak quenched bond randomness
converts the first-order phase transition of the TrSAFM to a
second-order transition. Thus, our finding strongly supports the
theoretical prediction of Refs.~\cite{aizenman-89,hui-89} that
quenched bond randomness always affects first-order phase
transitions by conversion to second-order phase transitions, for
infinitesimal randomness in $d=2$. The emerging, under random
bonds, continuous transition shows a strongly saturating specific
heat behavior very similar to the one found in our investigation
of the corresponding random-bond version of the square
SAF model~\cite{fytas-08a}. It also resembles the recently studied
analogous case of the induced second-order phase transition of the
originally first-order transition of the 2d BC
model~\cite{malakis-09}. These similarities point toward the
delicate nature of the corresponding pure system's phase
transitions, due to the competition of the microscopic nearest-
and next-nearest-neighbor exchange interactions, or the
competition between nearest-neighbor exchange interactions and the
crystal field in the case of the 2d BC model. It appears that such
competitions are very sensitive to the introduction of quenched
bond randomness.

We found that, the emerging second-order transition belongs to a
new distinctive universality class with the following critical
exponents $\nu=1.135(11)$, $\gamma/\nu=1.744(9)\approx 1.75$, and
$\beta/\nu\approx 0.124(8)\approx 0.125$. The numerical evidence
illustrated in figures~\ref{fig:4} and \ref{fig:5} clearly show
that the present model is out of the realm of a double logarithmic
behavior predicted theoretically and numerically verified for the
case of the marginal 2d
RBIM~\cite{fytas-08a,dotsenko-81,shalaev-84,shankar-87,ludwig-87b,wang-90,reis-96,ballesteros-97,selke-98,mazzeo-99}.
Our conclusion, is in general agreement with the predictions of a
distinctive universality class for the emerging second-order
transitions~\cite{falicov-96,ozcelik-08}, and the value
$\nu=1.135(11)$ of the correlation length's exponent is in
agreement with the inequality $\nu\geq 2/d$ for disordered systems
of Chayes et al.~\cite{chayes-86}.

The estimated magnetic critical exponents support an extensive but
weak universality as a wide variety of 2d systems
without~\cite{suzuki-74,gunton-75} and
with~\cite{fytas-08a,malakis-09} quenched disorder. However, since
in the case of the RBPM this property has been clearly ruled out
by the study of Cardy and Jacobsen~\cite{cardy-97}, it will be
very interesting to undertake a more extensive numerical study of
both square and triangular SAF models. Following
Picco~\cite{picco-98}, such a study may be extended to several
disorder strengths in order to observe the FSS behavior for the
disorder strength regime which is close to the random fixed point.

\begin{ack}
The authors would like to thank Prof. A.N. Berker for useful
discussions. This research was supported by the Special Account
for Research Grants of the University of Athens under Grant Nos.
70/4/4071. N.G. Fytas acknowledges financial support by the
Alexander S. Onassis Public Benefit Foundation.
\end{ack}

{}

\end{document}